\documentclass[prl,twocolumn,showpacs,preprintnumbers,amsmath,amssymb]{revtex4}

\usepackage{graphicx}
\usepackage{dcolumn}
\usepackage{bm}
\graphicspath{{./figures/}}
\usepackage{color}

\begin{document}

\preprint{arXiv:quant-ph/0506076 v1}

\title{Maser Oscillation in a Whispering-Gallery-Mode Microwave Resonator}

\author{P.-Y. Bourgeois, N. Bazin, Y. Kersal\'{e}, and V. Giordano}%
\email{giordano@lpmo.edu} \affiliation{Institut FEMTO--ST, Dpt.
LPMO, UMR 6174 CNRS-Universit\'{e} de Franche-Comt\'{e} \\
32 av.~de l'Observatoire, 25044 Besan\c{c}on Cedex, France}%
\author{M. E. Tobar}%
\email{mike@physics.uwa.edu}
\affiliation{School of Physics, University of Western Australia, Crawley 6009, WA, Australia}
\author{M. Oxborrow}%
\email{mo@npl.co.uk}%
\affiliation{National Physical Laboratory, Queens Road, Teddington, Middlesex TW11 OLW, UK}%

\date{\today}

\begin{abstract}
We report the first observation of above-threshold maser oscillation in a
whispering-gallery(WG)-mode resonator, whose
quasi-transverse-magnetic, 17$^{\rm{th}}$-azimuthal-order
WG mode, at a frequency of approx.~12.038~GHz, with
a loaded \emph{Q} of several hundred million, is supported on
a cylinder of mono-crystalline sapphire.
An electron spin resonance (ESR) associated with
Fe$^{3+}$ ions, that are substitutively included within the
sapphire at a concentration of a few parts per billion,
coincides in frequency with that of the (considerably narrower) WG mode.
By applying a c.w.~`pump' to the resonator at a frequency of approx.~31.34~GHz,
with no applied d.c.~magnetic field, the  WG (`signal') mode  is
energized through a three-level maser scheme.
Preliminary measurements demonstrate a frequency stability (Allan deviation)
of a few times $10^{-14}$ for sampling intervals up to 100~s.
\end{abstract}

\pacs{06.30.Ft,42.65.Pc,76.30.-v,76.30.Fc,84.40.Az,84.40.Ik}

\maketitle


Short-term ($<$100~s) fractional frequency stabilities better than $1\times 10^{-14}$ have
only ever been achieved at microwave frequencies with oscillators incorporating
cryogenic ($<$10~K) electromagnetic resonators exhibiting \emph{Q}~values in
excess of 100 million\cite{mann00_ultrastable}.
Such oscillators have been used successfully as `flywheels' for cold-atom
frequency standards~\citep{mann98}, as reference oscillators for (close-in)
phase-noise measurements\cite{dick92}, and in tests of fundamental physics (e.g.~Lorentz invariance\cite{wolf03}).

Two decades or so ago, Dick \emph{et al} developed a `superconducting cavity
maser oscillator' (SCMO)\cite{dick83,dick91}. It incorporated a cryogenic maser amplifier whose
ruby crystal was necessarily subjected to a d.c.~magnetic `bias' field. This amplifier
was intentionally separated from, yet electromagnetically coupled to, a high-\emph{Q} resonator
through an intermediate waveguide structure. The resonator took the form of a lead-coated sapphire
cylinder, maintained at a temperature near 1.6 K.

In recent years, the most actively studied resonators
have been those based on whispering-gallery (WG) modes, supported
on uncoated sapphire cylinders or rings that are
immediately surrounded by free space, where the cylinder's diameter-to-height
ratio is greater than unity, and where the resonator is generally
maintained at a temperature above 4.2~K\cite{chang00,dick98,uffc04_open_cavity}.
Here, the  bulk of the WG mode's field energy resides just within
the curved outer cylindrical wall of the sapphire monocrystal.
A Pound-stabilized-loop oscillator (PSLO)\cite{giles89} is
built around the WG-mode resonator, with the oscillator's sustaining
amplifier and phase modulator(s) located outside of the cryostat.
A PSLO is thus a spatially extended system; two microwave lines, each typically $>$1~m
in length, join the cryogenic resonator and room-temperature electronics
together in a loop. Moreover, to achieve stabilities at the $1\times 10^{-14}$ level,
additional circuits supporting the control of the resonator's temperature and
received microwave power\cite{luiten95} are required.

In all of the oscillators so far described, the electromagnetic resonator functions
as a purely passive, linear device (except potentially for a slight power-dependent
frequency shift).
In contrast, we report here the observation of continuous, above-threshold
maser\cite{siegman64} oscillation in an \emph{active} resonator. Here, amplification
is achieved through the interaction between a whispering-gallery mode and a collection
of ($\sim$$10^{15}$) paramagnetic ions that exhibit an electron spin resonance (ESR).
These ions are located, in space, within the WG mode's field profile and the WG mode
is located, in frequency, within the ESR's lineshape.
Compared to Pound-stabilized loop oscillators, or even Dick et al's SCMO, our
incorporation of maser gain within the oscillator's frequency-determining element
represents a fundamentally different approach.

Our whispering-gallery(-mode) maser oscillator, henceforth `WGMO', may be regarded as
a free-running loop oscillator, whose loop is the (closed) path taken by its
WG (signal) mode through space, and whose amplifier is continuously distributed around
this loop/mode. Some immediately apparent advantages are:
(i) the rigidity and compactness of the all-sapphire oscillator loop enables its electromagnetic
length (hence the WGMO's frequency --as determined by the Barkhausen condition) to be kept extremely constant;
(ii) unlike a ruby maser, no d.c. magnetic bias field need be applied;
(iii) compared to a PSLO or even the SCMO, the WGMO comprises
fewer essential components, \emph{viz.} just the sapphire cylinder and its
associated electromagnetic pump- and signal-mode couplers; there is no Pound frequency
servo; there are no cables or coupling structures between a spatially separated amplifier
and resonator;
(iv) moreover, the adjustment of the electromagnetic couplings to the maser oscillator's
signal and pump modes are, in contrast to the equivalent adjustments required for
optimizing a PSLO, far less critical.

\begin{figure}[h]
\centering
\includegraphics[width=0.50\columnwidth]{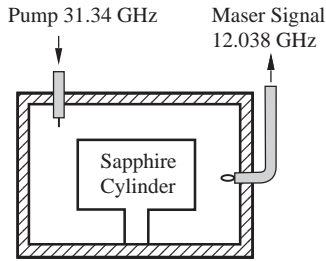}
\caption{\footnotesize \it{Principle of the Fe$^{3+}$ WG-mode maser oscillator.~~~~~~~~~~~~}} \label{anatomy}
\end{figure}

Half a century or so ago, the study of electron spin resonance in solids
led to the development of solid-state
masers~\cite{bloembergen56,siegman64} as sufficiently wideband, extremely
low-noise amplifiers for applications in satellite communications and radio astronomy.
Sapphire crystals deliberately doped with Fe$^{3+}$~ions, as opposed to Cr$^{3+}$,
were studied by a few groups\cite{kornienko59,bogle59_paramagnetic};
even a few maser amplifiers based on the former species were demonstrated\cite{king59,friedman63}.

Our electromagnetic resonator (see fig.~\ref{anatomy})
contains a monocrystal of HEMEX-grade\cite{crystalsystems05}
sapphire that comprises a main cylinder, 50~mm in diameter and 30~mm high, with a smaller,
coaxially adjoining cylinder (its `spindle') for support;
this monocrystal is mounted coaxially within a cylindrical copper cavity,
whose interior walls are silver-plated.
The monocrystal can support various whispering-gallery modes but only two of them,
both quasi-transverse-magnetic ($WGH$) in character, are presently relevant:
(i) a fundamental (i.e.~with no axial or radial
nodes) 17$^{\rm{th}}$-azimuthal-order $WGH$ mode,
at approx.~12.038 GHz and
(ii) a different, as yet unidentified $WGH$ mode of considerably
higher azimuthal-order at approx.~31.339 GHz. [Both of these frequencies refer to near-4.2~K operation.]
These two WG modes shall henceforth be referred to as the `signal' and `pump' modes, respectively.
The former is excited by an appropriately positioned and oriented loop probe
(sensitive to the magnetic field's azimuthal component), the latter by a stub antenna (sensitive to
the electric field's axial component).
The surrounding cavity is mounted within a vacuum can on the end
of a cryogenic insert, which is loaded into a liquid-helium dewar.
Microwave transmission lines, each comprising several
lengths of semi-rigid RG-405 coaxial cable joined by SMA connectors and
feedthroughs, connect each of the active resonator's two probes
to terminals on the insert's top plate.

Though our sapphire monocrystal was not intentionally doped,
ferric iron ions (Fe$^{3+}$) lie within it as residual impurities,
substituting for Al. Estimates in the literature for the
concentration of iron (presumably as Fe$^{3+}$) in samples of an unspecified grade of
nominally undoped HEM-grown sapphire~\cite{schmid73}, and in dielectric
resonators made from the HEMEX grade of the same~\cite{luiten96},
differ by orders of magnitude.
By model-fitting to both (a) the observed anomalous bistability
of the signal mode when directly driven (this phenomenon is
analogous to that of `optical bistability', as exhibited by a
saturable absorber \cite{lugiato84,oxborrow_unpub04}), and (b) the frequency shifts of different
X-band $WGH$ modes upon saturating the ESR at $\sim$12.04 GHz that is associated with the (paramagnetic)
Fe$^{3+}$ dopant (by driving the signal mode, that lies coincident with this ESR, sufficiently hard),
our own measurements indicate an effective substitutional concentration for Fe$^{3+}$ in HEMEX sapphire
of around a few parts per billion.

\begin{figure}[h]
\centering
\includegraphics[width=0.95\columnwidth]{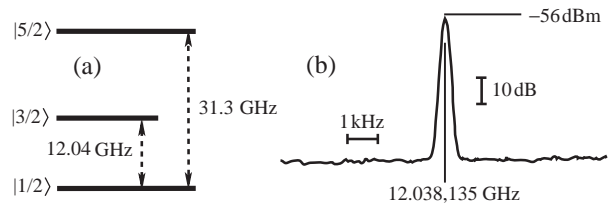}
\caption{\footnotesize (a) \it{Energy-level diagram for $Fe^{3+}$ in sapphire at
zero applied magnetic field and} \rm (b) \it{the unamplified maser signal as observed on
a spectrum analyzer (resolution bandwidth = {\rm 100~Hz}).}} \label{levels}
\end{figure}

The Fe$^{3+}$ ion's three paramagnetic energy levels at zero d.c.~magnetic
field are represented in fig.~\ref{levels}(a). Each of these is in fact a
degenerate Kramers doublet~\cite{king59,orton68}; the $S_z = \pm 5/2$
and $\mp 1/2$ spin states are furthermore mixed slightly together.
Transitions between these three levels are (thus) all allowed, though the
level-crossing pump transition is rather weak\cite{bogle59_paramagnetic};
their linewidths are all expected to be a few or several tens of~MHz.
Our WGMO exploits Bloembergen's classic
three-level scheme\cite{bloembergen56}, but at zero applied d.c.~magnetic field;
$\vert 1/2 \rangle \leftrightarrow \vert 3/2 \rangle$ transitions are stimulated
by the resonator's signal mode, whose frequency lies near the center of the
signal transitions' lineshape; similarly, $\vert 1/2 \rangle \leftrightarrow \vert 5/2 \rangle$
transitions are stimulated by the resonator's pump mode.

\begin{figure}[h]
\centering
\includegraphics[width=0.90\columnwidth]{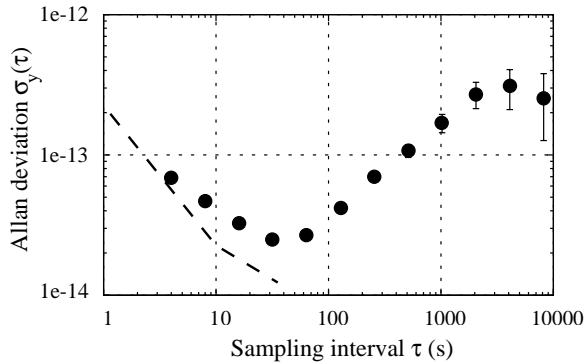}
\caption{\footnotesize \it{Frequency stability of WGMO;
the dashed line indicates the stability of the measurement's frequency reference.}}
\label{stab_maser4}
\end{figure}

The latter was excited, via its corresponding transmission line with terminating stub antenna,
by the (c.w.) output of an Agilent E8254A microwave frequency synthesizer. When this pump
synthesizer was set to a frequency of 31.339 GHz, and an output power level of 2~dBm,
a $-56$~dBm signal at approx 12.038135~GHz on the other transmission line, as connected to the
resonator's loop probe, could be detected at the insert's top plate --see fig.~\ref{levels}(b).
This signal, caused by maser oscillation on the $WGH_{17,0,0}$ signal mode, was
amplified by 70~dB then mixed (with a doubly balanced mixer) against
the signal from a second microwave synthesizer (Wiltron 69137A) referenced
to a commercial hydrogen maser. The resulting  beat-note (approx.~91~kHz in frequency)
was sent to a high-resolution frequency counter (HP~53132A).
By slowly increasing the resonator's temperature whilst monitoring this counter,
we observed the WGMO's signal frequency to turn over (a maximum) at a temperature of
approx.~7.939~K. The resonator's temperature was then stabilized at this turn-over
and the beat-note frequency measured against time. The corresponding
fractional-frequency Allan deviation\cite{howe81} was subsequently computed,
with the result shown in fig.~\ref{stab_maser4}.

\noindent The stability of the reference synthesizer limited the measurement's resolution
for sampling intervals $\tau< 20$~s. Nevertheless, a minimum in the Allan deviation
of $2.5\times 10^{-14}$ at 40~s was obtained.

We point out that the RG-405/SMA-based transmission line for conveying the Ka-band
pump down through the insert was not designed or tested for operation above 18~GHz.
For lack of appropriate equipment, the power reflected back from the pump probe,
hence the coupling to the pump mode, could not be quantified.
Assuming a spin-lattice relaxation time (`$T_1$') of a few ms\cite{bogle59_paramagnetic},
the measured $-56$~dBm level of maser signal power is consistent with the
parts-per-billion concentration of Fe$^{3+}$ ions inferred from our other
measurements (mentioned above) on the same sapphire monocrystal;
the $\vert 1/2 \rangle \leftrightarrow \vert 5/2 \rangle$
transition was not fully saturated at the 2~dBm level of applied
pump power used.

The origin of the long-term degradation in the frequency stability
has yet to be determined. No microwave isolators (for either X-band
or Ka-band) were placed in the transmission lines between the resonator and
the top plate within the insert; shifts in the VSWRs in sections of
these lines (caused by changes in the cryostat's temperature profile due
to boil-off of liquid helium) could thus
have significantly `pulled' the maser oscillator's frequency.
We now plan to evaluate the WGMO's sensitivities to pertinent
experimental variables including the pump power and frequency,
the signal- and pump-mode couplings, the loading and VSWR along
of the pump and signal transmission lines, as well as the
ambient and/or a deliberately applied magnetic field.


\bibliography{apl_2005_maser_arxiv_mo4} 

\end{document}